\documentclass[final,12pt]{elsarticle}

\usepackage{amssymb}
\usepackage{amsmath}
\usepackage{deluxetable}
\usepackage{caption}
\usepackage{multirow}
\usepackage{todonotes}
\usepackage{tablefootnote}
\usepackage{orcidlink}
\usepackage{lineno}
\usepackage{natbib}
\usepackage{subcaption}
\usepackage{caption}

\journal{NIM A}

\begin{document}

\begin{frontmatter}



\title{Proton Radiation Damage and Annealing of COSI p-type Cross-strip HPGe Detectors}


\author[a]{Sophia E. Haight\orcidlink{0000-0003-3469-7072}}
\author[a,b]{Steven E. Boggs\orcidlink{0000-0001-9567-4224}}
\author[a]{Gabriel Brewster\orcidlink{0009-0005-2936-8516}}
\author[a]{Sean N. Pike\orcidlink{0000-0002-8403-0041}}
\author[a]{Jarred M. Roberts\orcidlink{0000-0002-7660-2740}}
\author[c]{Albert Y. Shih\orcidlink{0000-0001-6874-2594}}
\author[d]{Joanna M. Szornel\orcidlink{0009-0002-1776-5475}}
\author[b]{John A. Tomsick\orcidlink{0000-0001-5506-9855}}
\author[a]{Aravind B. Valluvan\orcidlink{0000-0002-4264-6112}}
\author[b]{Andreas Zoglauer\orcidlink{0000-0001-9067-3150}}


\affiliation[a]{organization={Department of Astronomy \& Astrophysics, University of California, San Diego},
            addressline={9500 Gilman Drive}, 
            city={La Jolla},
            state={CA},
            postcode={92093}, 
            country={USA}}
            
\affiliation[b]{organization={Space Sciences Laboratory, University of California, Berkeley},
            addressline={7 Gauss Way}, 
            city={Berkeley},
            state={CA},
            postcode={94720}, 
            country={USA}}
\affiliation[c]{organization={NASA Goddard Space Flight Center},
city={Greenbelt}, state={MD}, postcode={20771}, country={USA}}

\affiliation[d]{organization ={Lawrence Berkeley National Laboratory},
    addressline={1 Cyclotron Rd}, 
    city ={Berkeley}, 
    state ={CA}, 
    postcode ={94720}, 
    country={USA}}

\begin{abstract}
In order to understand the effects of a space radiation environment on cross-strip germanium detectors, we investigated the effects of high-energy proton damage on a COSI detector and the capabilities of high-temperature annealing in repairing detector spectral resolution. We irradiated a COSI-balloon cross-strip high-purity germanium (HPGe) detector with 150\,MeV protons resulting in a net fluence of $4.95\times10^8$ p$^+$/cm$^2$ and corresponding to $\sim$\,10 years in COSI's space radiation environment. We repaired the resulting degradation in spectral resolution through a series of high-temperature anneals to obtain a final FWHM of 4.08\,keV, within 37\,$\%$ of its preradiation value (2.98\,keV FWHM). We characterized the repair of charge traps with time spent under high-temperature anneal to inform an annealing procedure for long-term maintenance of COSI's spectral resolution. 
\end{abstract}


\end{frontmatter}


\section{Introduction}
\label{sec:introduction}
\subsection{Background}
\label{subsec:background}
The Compton Spectrometer and Imager (COSI) is a NASA Small Explorer (SMEX) satellite mission currently under development for launch in 2027 \cite{tomsick2023}. COSI is a wide-field gamma-ray telescope designed to provide spectroscopy, imaging, and polarimetry of astrophysical sources in the bandpass 0.2-5 MeV. The heart of the COSI instrument is an array of 16 cross-strip germanium detectors (GeDs) providing excellent energy resolution (0.5\% at 0.662\,MeV for single-strip events in \citet{Beechert2022}) to aid in two of the primary mission goals. First, to uncover the origin of Galactic positrons through imaging spectroscopy of the annihilation emission at 0.511\,MeV. And second, to reveal Galactic element formation through study of gamma-ray emission lines produced through radioactive decay of recently synthesized elements. To fulfill these science goals, COSI must maintain its spectral performance throughout its primary mission.  

\indent The spectral resolution of GeDs in space degrade due to increased charge trapping induced by radiation damage from ionizing particles, most significantly high-energy protons. High-energy protons damage the crystal lattice structure of the detectors and produce negatively ionized deformations which trap holes as they drift in the applied bias voltage towards the signal electrodes \cite{Darken1980}. Trapping degrades spectral resolution of the hole-collection signal, characterized by the broadening of the photopeaks, as measured by the full width at half maximum (FWHM), and increased low-energy tailing of spectral lines. This degrades the ability to determine the energy of the incoming gamma-rays.


\indent Similar radiation damage \text{eff}ects have been observed by germanium spectrometers INTEGRAL/SPI \cite{Vedrenne2003} and RHESSI \cite{Lin2002} during their time in orbit. Both instrument teams implemented a series of high-temperature anneals at $\sim$\, 100$^{\circ}$C to repair the radiation damage and restore the spectral resolution. However, annealing at these temperatures in orbit requires careful planning and design to avoid permanent damage to any instrument. The SPI telescope is composed of 19 coaxial GeDs launched into a highly eccentric orbit (perigee of 9000 km and apogee of 154600 km). After noting a significant degradation in energy resolution after 2 years in orbit, SPI began completing periodic high-temperature anneals to repair spectral resolution and prevent permanent damage from radiation \cite{Lonjou2005}. SPI currently anneals at 100$^{\circ}$C for 14 days every 6 months, observing an improvement of within 10-15\,$\%$ to their spectral resolution between each anneal \cite{Diehl2018}. However, they suffered the loss of 2 GeDs \cite{Lonjou2005}. In low earth orbit (LEO) at 38$^\circ$ inclination, RHESSI similarly experienced significant losses in resolution which warranted an annealing procedure. Nine segmented coaxial n-type GeDs underwent 5 anneals at 100$^{\circ}$C over 9 years of operation, each anneal lasting for approximately 1 week \cite{Dennis2022}. Although the length of each anneal was minimized to mitigate the risk of fusing the spectrometer's segmented germanium crystals at the lithium-drifted contacts, multiple RHESSI detectors became permanently unsegmented after the fourth anneal \cite{Shih2024}.


\indent COSI will be launched into an equatorial LEO (530 km, 0$^\circ$ inclination). In this orbit, COSI radiation damage will be dominated by trapped protons. The COSI GeDs will be enclosed by an aluminum cryostat which blocks protons $<$\,20 MeV, and an active Bismuth Germinate (BGO) shield in a shallow well configuration, shielding the bottom and sides of the GeD array from atmospheric gamma-rays, and absorbing the majority of protons $<$\,150 MeV which are incident from directions outside of the main field of view of the instrument. The main contribution of protons $>$\,20 MeV will accumulate from the times during which COSI's orbit passes through the trapped proton fluence in the edge of the South Atlantic Anomaly (SAA). Two simulated space radiation environment models AP8 \cite{Sawyer_Vette_1976} and AP9 \cite{Ginet2013} provide estimates of the fluence originating from the SAA. Their predictions differ by two orders of magnitude, while the measured proton flux from X-ray telescope BeppoSAX (in LEO at $4^{\circ}$ inclination) falls in between these models \cite{Ripa2020}. For our study, we adopt the conservative upper limit set by AP9 which estimates an \text{eff}ective proton fluence exposure of $1.1 \times 10^8$  p$^+$/cm$^2$ over COSI’s primary 2-year mission (95\,$\%$ confidence level). The fluences studied in this work are based on a previous predicted orbit for COSI which would have resulted in longer paths through the SAA, higher proton fluences, and greater damage to GeDs. In 2024, COSI's orbit was confirmed at a near equatorial inclination, resulting in a lower proton fluence. As a result, the net fluence studied in this work corresponds roughly to an extended 10 year mission at COSI's confirmed orbit. 

\indent Neutron radiation induced damage to GeDs has been studied extensively \cite{Darken1980, Fourches_and_Walter, Fourches1991,Kandel1999}. However, relatively few studies of proton irradiation on detector performance provide insight for this work. Initial study of proton radiation was carried out on planar p-type GeDs with 6 GeV protons providing fluences ranging (0.73 - 5.7)$ \times 10^8$ p$^+$/cm$^2$ \cite{Pehl1978}. The effects of proton irradiation included prominent low-energy tailing and an increase in FWHM which was approximately linear with fluence. The measurements from these studies concluded that GeV protons induce damage at lower fluences than irradiation from MeV neutrons. Furthermore, sharp peaked distributions post proton damage suggest that the damaging mechanism for protons is distinct from that of neutrons. 1.5 GeV proton irradiation of p- and n-type coaxial GeDs with net fluences on each detector within the range of $(1.18-1.34) \times 10^8$ p$^+$/cm$^2$ likewise produced significant increase in FWHM and tailing at low energies \cite{Koenen1994}. Testing three detectors in parallel, while holding each at a different temperature during irradiation, revealed a greater degradation in spectral resolution at higher detector temperatures within a range of 90\,K to 130\,K.

Our objectives for this work are to characterize the effects of radiation damage on COSI's spectral performance over the primary and extended missions, and to formulate and test an annealing procedure to repair said damage and restore COSI's spectral resolution to its baseline value. Through these damage and annealing procedures, we track the spectral resolution by measuring the effective FWHM of the spectral line from decaying isotope Cesium 137 (Cs-137), emitting gamma-rays at 0.662 MeV. We utilize numerical charge transport simulations, including a model of trapping \cite{Boggs_Pike_2023} to relate the density of charge traps within the detector volume to the resulting degradation in spectral resolution, and ultimately track how trapping defects decrease with time spent annealing at high temperatures.

\section{Carrier Transport and Trapping}
\label{sec:trapping}

\subsection{Primary and Secondary Traps}

In this study, we attempt to understand and separate the effects on spectral resolution of (i) primary damage caused by irradiation and (ii) secondary damage caused by cycling detectors to room temperature. We draw on previous studies to lay out a basic model of charge carrier trapping after irradiation and cycling detectors to room temperature \cite{Boggs_Pike_2023}. 

The high purity germanium used in COSI detectors has a crystalline lattice structure in which charge carriers can travel, freed by incoming gamma-rays or other ionizing radiation. Primary defects are formed from irradiating the germanium with high-energy particles. These particles scatter either individual atoms or a cascade of neighboring atoms out of their position in the lattice, resulting in lattice ‘vacancies’ and atomic displacements called interstitials. Vacancies can be either individual \cite{Brown1953} or form ‘disordered regions’ in a stream of successive atomic displacements \cite{Darken1980}. Both types of vacancies act as traps for charge carriers, preferentially for holes.  


Secondary defects form as the result of interstitial motion which takes place when the detectors are cycled to room temperature. \citet{Fourches_and_Walter} used deep-level transient spectroscopy (DLTS) to track signatures of traps, and concluded that secondary damage is caused by the migration of individual vacancies out of the primary disordered regions, potentially combining to form divacancy sites. These sites also act as hole traps. Confirmed in previous works \cite{Thomas1993, Kandel1999} and in our experience temperature cycling this detector, secondary defects do not form when undamaged germanium is cycled to room temperature, only when primary defects are already present. The formation of secondary defects is important to understanding the potential impacts of radiation damage as previous studies observe a more significant degradation in resolution from secondary defects than primary defects \cite{Fourches_and_Walter, Kandel1999}. 


\subsection{Trapping Parameters}
\label{subsec:trappingparams}

\indent On a basic level, charge trapping is determined by the density of charge traps in the material, $n$, and the cross section of the traps, $\sigma$.  The total trap density will be determined by an intrinsic trap density, originating from impurities introduced in the fabrication process, and the trap density induced by external damage -- in this case irradiation and room temperature cycling. The trapping cross section $\sigma$ has been shown to vary by nearly two orders of magnitude for different trapping defects when measured with DLTS \cite{Fourches1991}. When a charge cloud $q_{0}$ travels a total path length $l$, the charge which has not been trapped at this point $q$ depends on a combination of $n$ and $\sigma$ such that: 

\begin{equation}
q=q_{0} e^{-l/[n\sigma]^{-1}}
\end{equation}

As derived in \citet{Boggs_Pike_2023}, the total path length is the vector combination of the net drift displacement and random thermal motions around the path of travel. Given the carrier drift velocity, $v_d$, and random thermal velocity, $v_{th}$ the total distance traveled over time $t$ is given by:

\begin{equation}
l = (v_{th}^2 + v_{d}^2 )^{1/2} t  
\end{equation}

\begin{figure}
    \centering
    \includegraphics[width=0.65\linewidth]{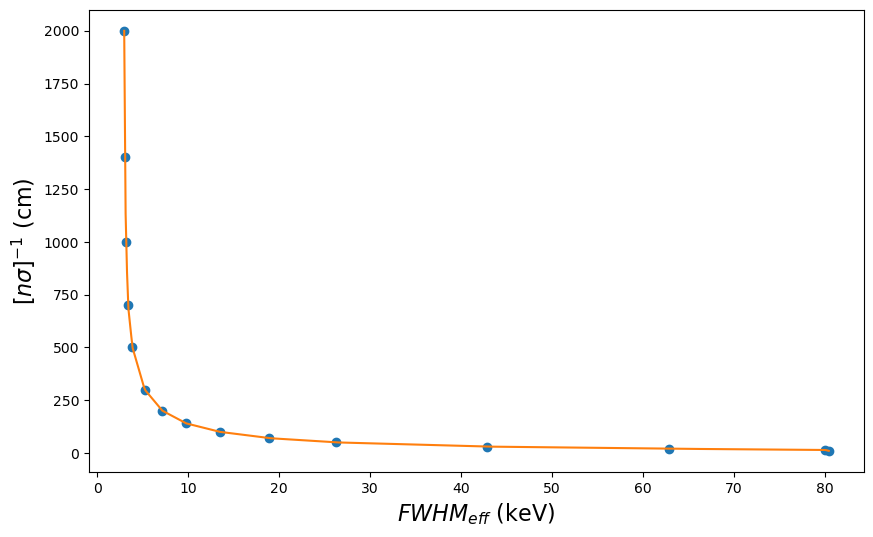}
    \caption{Linear interpolation between trapping products ($[n\sigma]^{-1}$) of simulated detectors and corresponding FWHM$_{\text{eff}}$ measured from simulated data. Data points are plotted in blue and the linear interpolation in orange. }
    \label{fig:linear_interp}
\end{figure}

We characterize trapping via the product $[n\sigma]^{-1}$, termed the `trapping product' and measured in units of cm. Under this definition, no charge trapping would correspond to an infinite trapping product. The lower the trapping product, the more trapping that occurs for a given path length $l$. Neither $n$ nor $\sigma$ can be measured directly from our detector resolution measurements. However, using simulations of our detectors for various trapping products, we can derive an empirical relation (Figure \ref{fig:linear_interp}) between $[n\sigma]^{-1}$ and the resulting spectral resolution, which we discuss further in Section \ref{subsec:simulations}. 

\section{Methods}
\label{sec:Methods}
\begin{figure}
    \centering
    \includegraphics[width=0.5\linewidth]{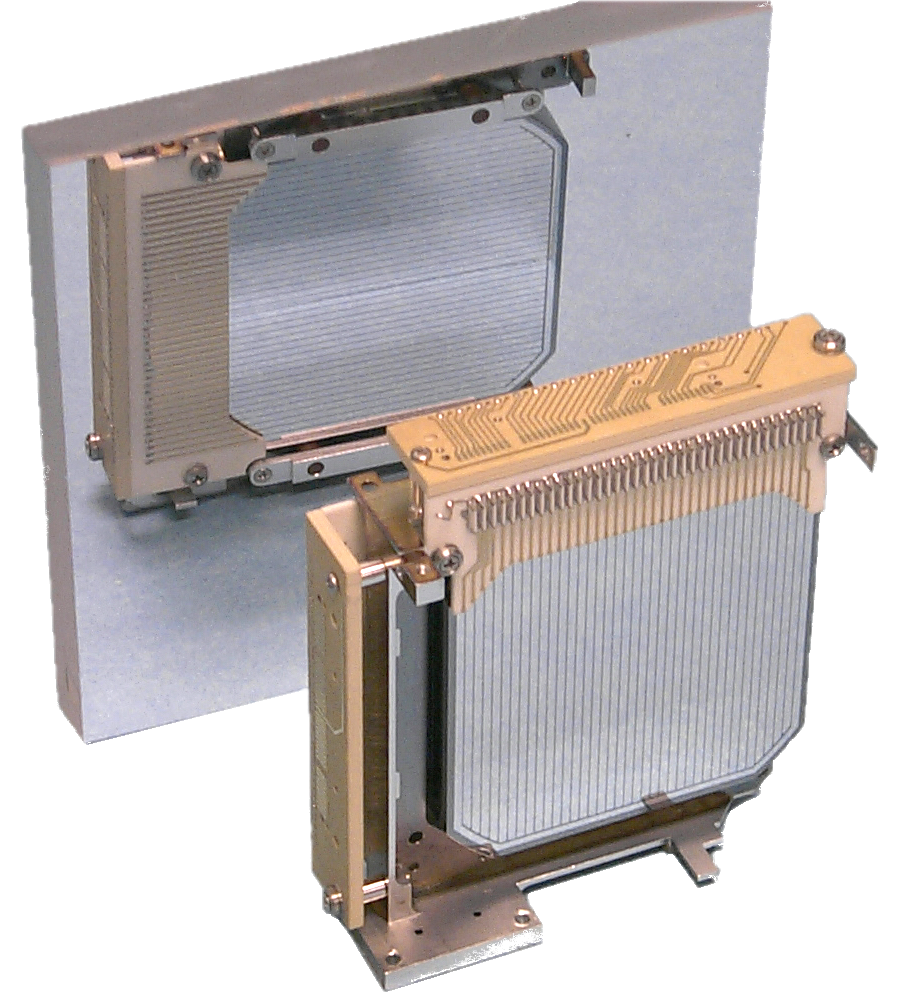}
    \caption{A spare COSI balloon detector with 37 electrode strips separated by high purity germanium crystals. A mirror shows the orthogonal strips on the opposite face of the detector.}
    \label{fig:GED}
\end{figure}

\begin{figure}
    \centering
    \includegraphics[width=0.7\linewidth]{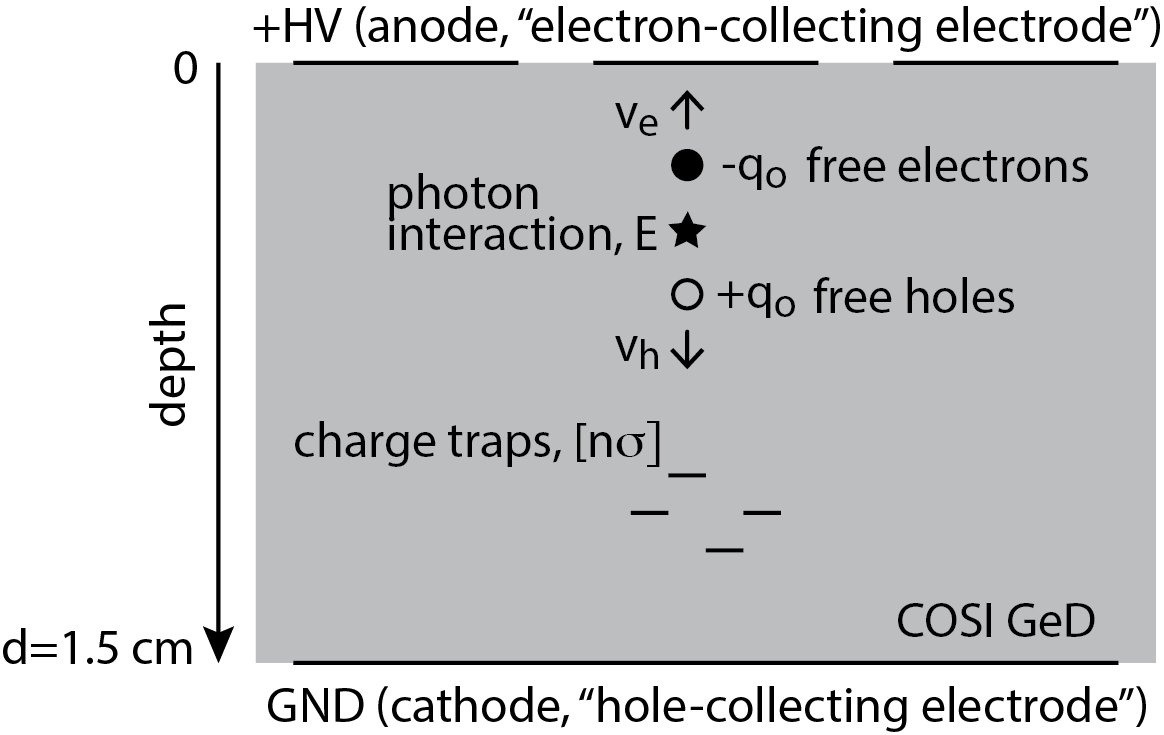}
    \caption{Charge transport in COSI GeDs. Electrons and holes drift at velocities $v_e$ and $v_h$ respectively in an applied voltage between two parallel electrode strips like those pictured in the full detector view in Figure \ref{fig:GED}.}
    \label{fig:trapping_model}
\end{figure}

For this study, we used a spare p-type high purity Germanium (HPGe) cross-strip detector from the COSI-Balloon mission \cite{Beechert2022}. The detector, shown in Figure \ref{fig:GED}, has dimensions of 8 cm x 8 cm x 1.5 cm and 37 electrode strips of 2-mm pitch on each face. This detector depletes at 400V, and we operate it at an applied bias of +600V within an evacuated cryostat cooled by liquid nitrogen to $\sim$\,80K. A gamma-ray interacting with the germanium crystal generates a cloud of electron-hole pairs from the material. Freed charges drift in the applied electric field, with the hole charge cloud collected on the LV (ground) electrode strip and the electron charge cloud on the HV (high voltage) electrode strip, as pictured in Figure \ref{fig:trapping_model}. Each electrode is connected to readout electronics which process the HV and LV signals separately; These are the exact readout electronics used on the COSI-Balloon payload as described in \citet{Beechert2022}. Events are then binned into energy spectra for analysis. To calibrate the detector and measure the spectral resolution prior to any radiation or annealing, we collected photopeak events from gamma-ray sources across a range of energies in the MeV band. \\
\indent In this work, we focus on spectral measurements of the 0.662\,MeV line from Cs-137. Due to non-gaussian and bimodal line profiles resulting from irradiation and temperature cycling the detector, we chose to characterize the detector resolution with an effective FWHM measurement. This FWHM$_{\text{eff}}$ is defined as $2.355 \times \sigma$ in terms of the standard deviation ($\sigma$) of counts in the Cs-137 peak which traditionally measures the variation of data points about their mean for gaussian and non-gaussian distributions alike. The choice to report resolution results in terms of FWHM${_\text{eff}}$ rather than $\sigma$ is for comparison with spectral resolution FWHM measurements in other works. We recognize that this non-standard FWHM${_\text{eff}}$ does not permit direct comparison to traditional FWHM measurements of non-gaussian distributions; however, it is well-defined and more easily measured than the traditional FWHM measurements for non-gaussian distributions. This FWHM${_\text{eff}}$ also has the advantage of converging with the traditional FWHM measurements for gaussian distributions with low background. The robustness of this measurement enables us to consistently track the spectral resolution across the variety of line profiles measured in this work. To determine the FWHM${_\text{eff}}$, we constrain peak counts and attempt to remove background counts from each of our distributions by considering energies higher than the energy at which three consecutive bins have counts below 15\% of the peak height and lower than 0.670\,MeV. We make an exception, fixing a lower bound at 0.475\,MeV, for data collected directly after room temperature cycling because the photopeak does not drop off to background levels and instead merges with the Compton backscatter continuum. Each distribution defining the FWHM$_{\text{eff}}$ for hole-collecting strips along with corresponding standard deviation ($\sigma$) measurements are shown in Figure \ref{fig:experimental_peaks}.

\subsection{Radiation Damage}
In order to replicate the proton fluence we might expect over COSI’s mission, we used the proton synchrotron at the James M Slater MD Proton Treatment and Research Center at Loma Linda University Medical Center. We irradiated our detector in two rounds with protons with an average energy of 150 MeV incident on the HV face of the detector. During the first round, the detector received a fluence of $2.00\times10^8$ p$^+$/cm$^2$. Then, after measuring the spectral resolution of the detector, we delivered another $2.95\times10^8$ p$^+$/cm$^2$, resulting in a net fluence of $4.95\times10^8$ p$^+$/cm$^2$. During irradiation, the detector was unbiased and kept under vacuum at operating temperature ( $\sim80$K). Our measurements of the spectral resolution after each round of radiation revealed a significant broadening in photopeaks of hole-collecting strips, characteristic of increased hole trapping due to radiation damage. After completing our post-irradiation measurements, we cycled the detector to room temperature for 336 hours while keeping it under vacuum.

\subsection{Annealing}
To repair the primary damage from the proton fluence, in addition to the secondary damage caused by cycling the detector to room temperature after irradiation, we proceeded with seven incremental anneals to characterize the progress towards restoring the spectral resolution. These anneals are listed in Table \ref{table}.

After each annealing procedure we returned the detector to its operating temperature and collected data from a Cs-137 source, tracking the spectral resolution through the resulting FWHM$_{\text{eff}}$ of the 0.662 MeV line. Radiation damage and annealing at room temperature produced non-gaussian line profiles including low-energy tailing, spectral line shifts, and bimodal distributions. To probe the effectiveness of annealing procedures at different temperatures, we completed anneals at $80^{\circ}$C and $100^{\circ}$C. The two anneals at $80^{\circ}$C were intended to characterize annealing at lower temperatures while, the 5 anneals at $100^{\circ}$C were intended to characterize the progress of the spectral resolution back towards its preradiation value. 

 \subsection{Numerical Simulations}
 \label{subsec:simulations}

In parallel to our measurements characterizing the effects of radiation and annealing on the spectral performance, we used numerical simulations to model the photopeak spectrum for a range of trapping products encompassing the anticipated levels of radiation damage. Our numerical charge transport simulations, including the model for charge trapping, are described and tested in \citet{Boggs_Pike_2023}. 

The measured FWHM$_{\text{eff}}$ of the photopeak has two components: FWHM$_{\text{eff},d}$ attributed to the trap density caused by proton damage, and FWHM$_{\text{eff},0}$ attributed to a combination of intrinsic trap density, statistical Fano noise, and noise from electronic readout. These components add in quadrature:

\begin{equation}
FWHM_{\text{eff}} = \sqrt{FWHM_{\text{eff},0}^2 + FWHM_{\text{eff},d}^2}
\end{equation}

For hole-collecting strips, we equated FWHM$_{\text{eff},0}$ to the measured preradiation FWHM$_{\text{eff},h}$ ($\sim$\, 2.98\,keV). 
We tabulated the FWHM$_{\text{eff}}$ of the Cs-137 photopeak produced by our numerical simulations against their respective trapping products and linearly interpolated between these data points to obtain a relationship between the two parameters. The interpolation and data points are plotted in Figure \ref{fig:linear_interp}. We used this interpolation to pair the measured FWHM$_{\text{eff}}$ produced at each stage of damage and annealing with the corresponding hole trapping products. 




\section{Results}
\label{sec:Results}

\begin{figure}
    \centering
    \includegraphics[width=1.0\linewidth]{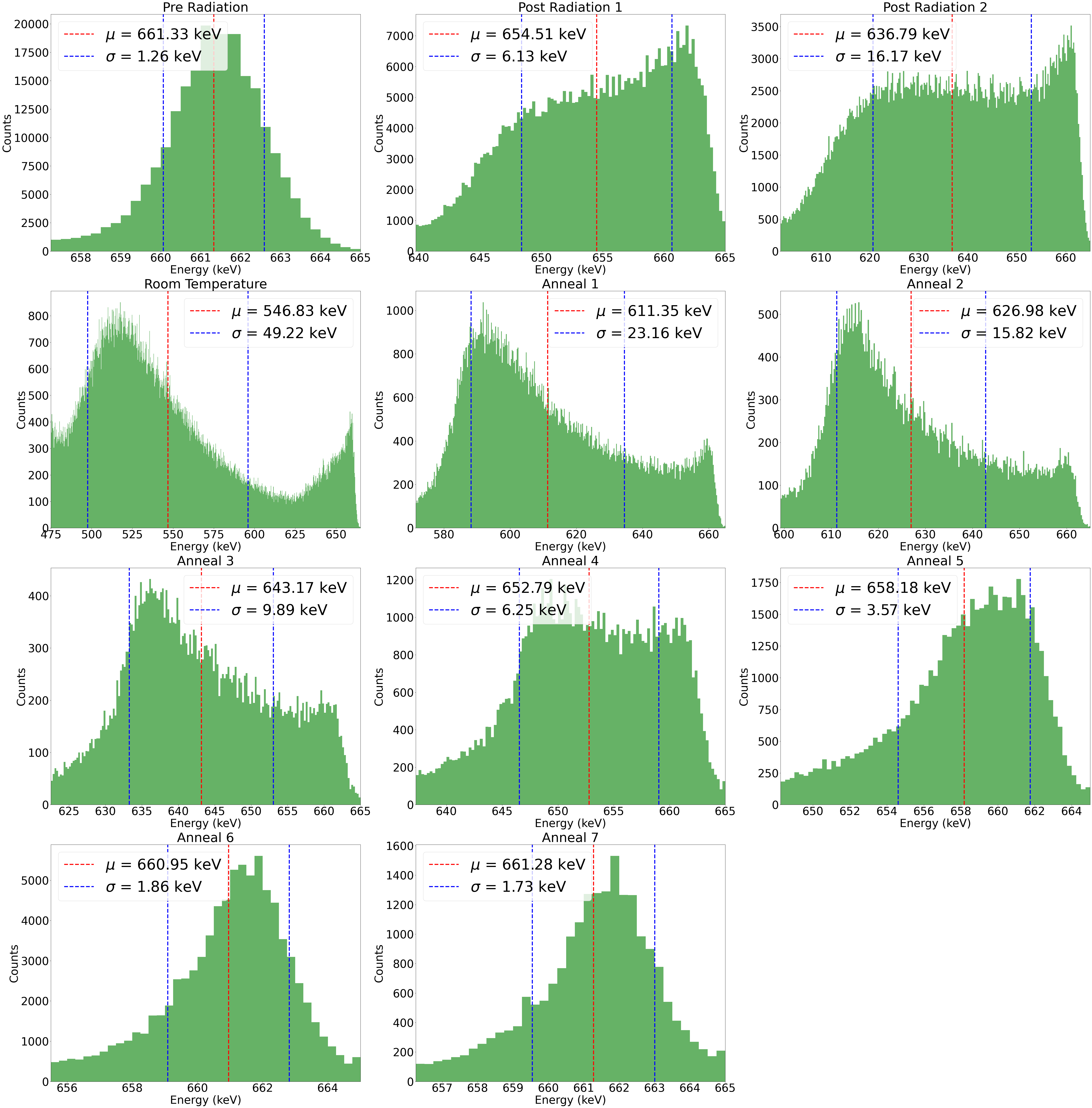}
    \caption{Spectra for hole-collecting strips for a Cs-137 source collected at each stage of irradiation and annealing. A red dashed line marks the mean ($\mu$) of the counts and two blue dashed lines mark one standard deviation ($\sigma$) in either direction. Corresponding FWHM$_{\text{eff}}$ values for each peak are listed in Table \ref{table}.}
    \label{fig:experimental_peaks}
\end{figure}

\begin{deluxetable}{ccccccc}
    \tablenum{1}
    \label{table}
    \tablecaption{Description of annealing and irradiation procedures along with the FWHM$_{\text{\text{eff}}}$ measured afterwards for electron and hole-collecting strips, FWHM$_{\text{eff}, e}$ and FWHM$_{\text{eff}, h}$ respectively. Columns list: (i) temperature the detector was annealed at (ii) duration of the anneal (iii) FWHM$_{\text{\text{eff}}}$ for hole-collecting strips, (iv) FWHM$_{\text{eff}}$ for hole-collecting strips where preradiation FWHM$_{\text{eff}}$ is subtracted in quadrature leaving only FWHM$_{\text{eff}}$ attributed to damage from radiation and room temperature annealing (described in Section \ref{subsec:simulations}), (v) hole trapping products calculated from numerical simulations outlined in Section \ref{subsec:simulations} (vi) FWHM$_{\text{eff}}$ for electron-collecting strips. }
    
    \tablewidth{0pt}
    \tabletypesize{\scriptsize} 
    \tablehead{
        & \colhead{Annealing Temp.} &
        \colhead{Annealing Time} &
        \colhead{FWHM$_{\text{eff}, h}$} &
        \colhead{FWHM$_{\text{eff}, d, h}$} &
        \colhead{$[n\sigma]_h^{-1}$} &
        \colhead{FWHM$_{\text{eff}, e}$} \\
        & \colhead{($^{\circ}$C)} &
        \colhead{(Hours)} &
        \colhead{(keV)} &
        \colhead{(keV)} &
        \colhead{(cm)} &
        \colhead{(keV)}
    }
    \startdata
    Preradiation & -- & -- & 2.98 & -- & 1786\tablenotemark{1} & 4.87 \\
    Radiation 1  & -- & -- & 14.4  & 14.1 & 95 & 4.3 \\
    Radiation 2  & -- & -- & 38.1    & 38.0 & 36 & 5.0 \\
    Room Temp    & $\sim 20$ & 336 & 115.9   & 115.9   & $<$10 & 7.1 \\
    Anneal 1     & 80  & 24  & 54.5    & 53.9 & 24 & 4.3 \\
    Anneal 2     & 80  & 48  & 37.2    & 37.1 & 37 & 3.8 \\
    Anneal 3     & 100 & 24  & 23.3    & 23.1 & 58 & 3.6 \\
    Anneal 4     & 100 & 48  & 15.7    & 14.4 & 93 & 3.9 \\
    Anneal 5     & 100 & 96  & 8.41   & 7.86  & 308 & 4.2 \\
    Anneal 6     & 100 & 192 & 4.39   & 3.22  & 971 & 4.1 \\
    Anneal 7     & 100 & 192 & 4.08  & 2.79  & 1313 & 4.1 \\
    \enddata

\tablenotetext{1}{Derived in \citet{Pike_2025}}
\end{deluxetable}

\subsection{Preradiation}
We measured the FWHM$_{\text{eff}}$ of the Cs-137 photopeak for hole and electron-collecting strips of 2.98\,keV and 4.87\,keV respectively, prior to any radiation damage or annealing. The broader resolution on the electron-collecting strips is due to HV-coupling and intrinsic electron trapping \cite{Pike2023}. The photopeak for hole-collecting strips is pictured in Figure \ref{fig:experimental_peaks}.

\subsection{Post Radiation}
\label{subsec:postrad}
After the first round of 150\,MeV proton irradiation at a fluence of $2.00\times10^8$ p$^+$/cm$^2$, we measure a peak with a 14.4\,keV FWHM$_{\text{eff}}$ for hole-collecting strips -- increased nearly by a factor of five from preradiation. After the second round of irradiation, resulting in a net fluence of $4.95\times10^8$ p$^+$/cm$^2$, our measured FWHM$_{\text{eff}}$ increased to 38.1\,keV, over 12 times its preradiation value. In terms of fluence, this reflects an increase of roughly 7.2\,keV per $10^{8}$ p$^+$/cm$^2$  after the first round of radiation and 7.7\,keV per $10^{8}$ p$^+$/cm$^2$  after both rounds of radiation. These measurements demonstrate a roughly linear relationship between detector resolution and net proton fluence. This trend is confirmed in \citet{Fourches_and_Walter} and \citet{Fourches1991} for p-type GeDs, and in \citet{Kandel1999} for SPI's n-type GeDs, all of which underwent neutron irradiation. Likewise, this is seen in \citet{Pehl1978} for p-type GeDs which underwent proton irradiation. We observe widened photopeaks and low-energy tailing (Figure \ref{fig:radiation}) characteristic of radiation damage. 


\subsection{Post Room Temperature Cycling}
\label{subsec:roomtemp}

\begin{figure}
    \centering
    \includegraphics[width=1.0\linewidth]{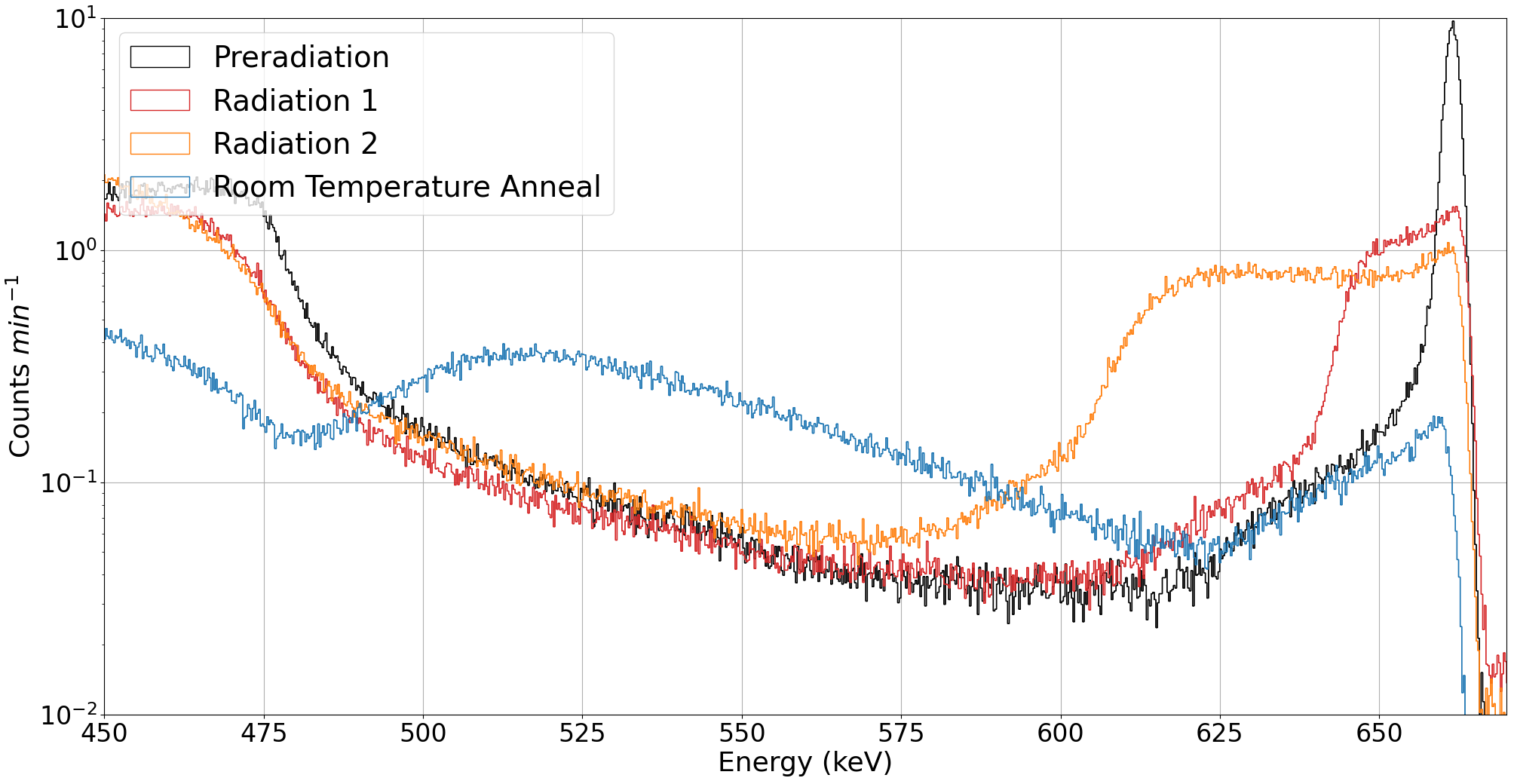}
    \caption{Counts per minute of hole-collecting strips for Cs-137 source plotted on a logarithmic scale. Counts collected: prior to radiation or annealing (black), after receiving $2.00\times10^8$ p$^+$/cm$^{2}$ (red), after receiving a total $4.95\times10^8$ p$^+$/cm$^{2}$ (orange), and after annealing at room temperature for 14 days (blue).}
    \label{fig:radiation}
\end{figure}

\begin{figure}
    \centering
    \includegraphics[width=1.0\linewidth]{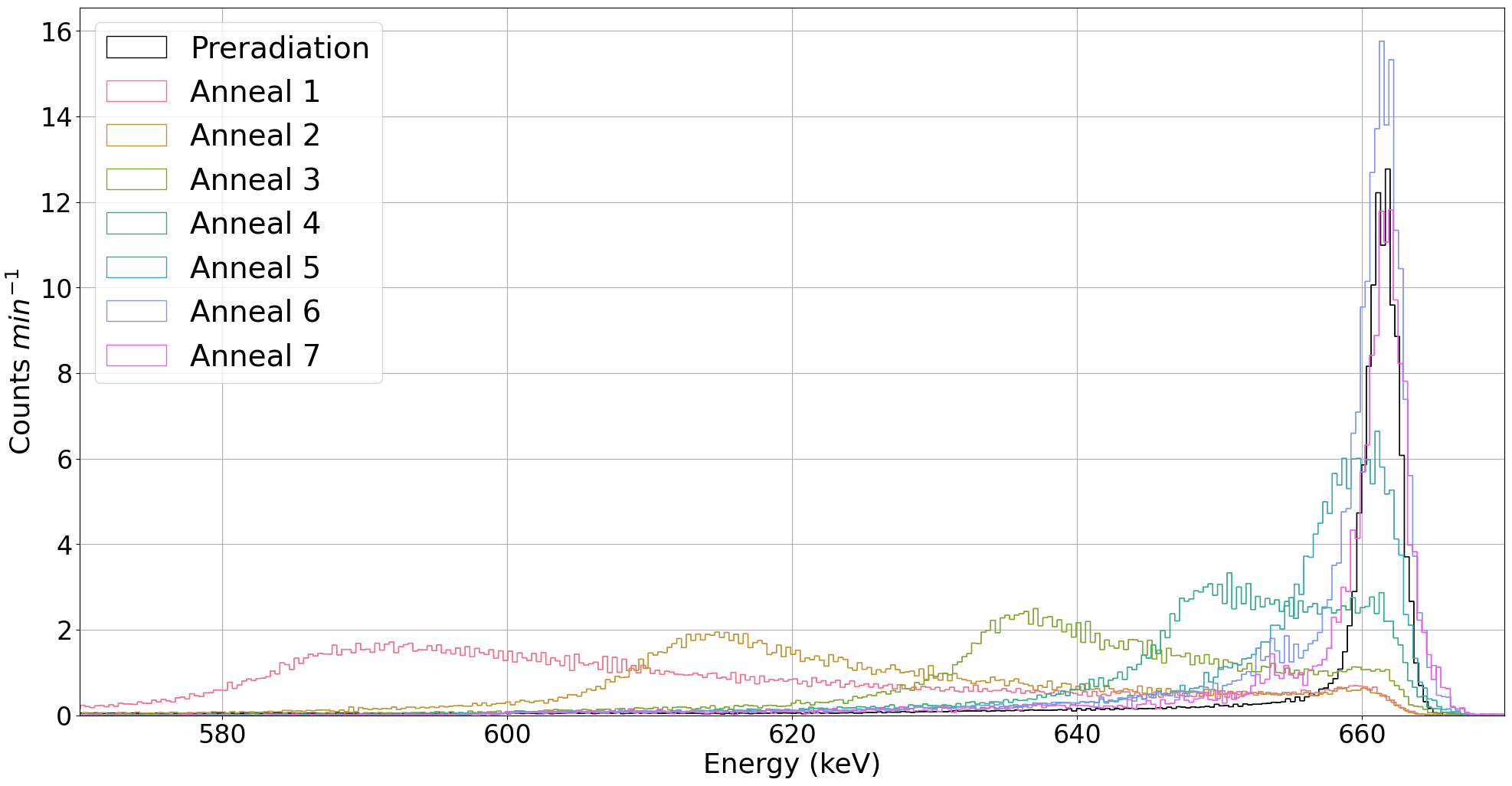}
    \caption{Counts per minute for hole-collecting strips for a Cs-137 source across incremental high-temperature anneals. Anneals and their respective FWHM$_{\text{eff}}$ measurements are listed in Table \ref{table}. }
    \label{fig:anneals}
\end{figure}

\begin{figure}
    \centering
    \includegraphics[width=1.0\linewidth]{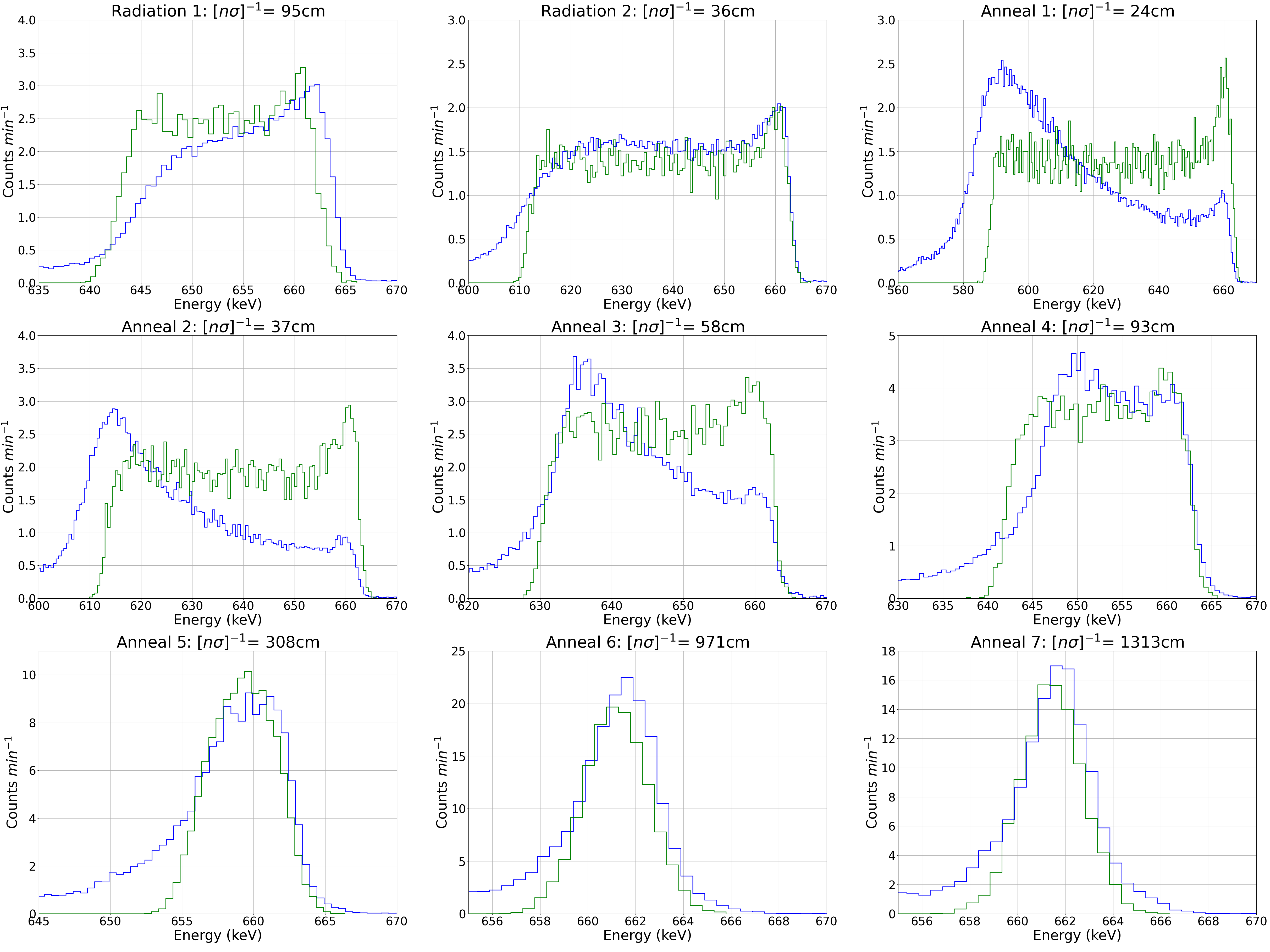}
    \caption{A comparison of line profiles of experimental photopeaks and photopeaks produced by numerical simulation at the same trapping product, plotted in blue and green, respectively. The height of the simulated photopeaks are arbitrarily scaled for comparison with experimental data.}
    \label{fig:simulatedpeaks}
 \end{figure}

Previous studies report detrimental effects on the spectral resolution of radiation-damaged detectors cycled to room temperature \cite{Pehl1978, Fourches1991, Peplowski2019, Kandel1999}. To understand the extent of this damage, we carried out our initial anneal by bringing the detector to $\sim 20^{\circ}$C for 336 hours. The photopeak we measured after this anneal, shown in Figure \ref{fig:radiation}, appears nearly bimodal and the detector was rendered unusable as a spectrometer. \\
\indent Furthermore, the line profile is unlike any of the profiles we observed after solely primary damage. As expected, a small peak appears at the expected 0.662\,MeV, however, the physical origin of the more prominent peak is unclear. It is downshifted to approximately 0.520\,MeV and merged with the Compton backscatter continuum we expect from Cs-137 around 0.48\,MeV. This downward shift, which can also be observed in photopeaks prior to anneal 4 (Figure \ref{fig:anneals}), is due to extreme trapping, demonstrating that the detector had a high trap density. However, in our charge transport models simulated with high trap densities/ low trapping products, we fail to observe this artifact. As demonstrated in Figure \ref{fig:simulatedpeaks}, simulated photopeaks with lower trapping products have FWHM$_{\text{eff}}$ measurements similar to those of their experimental counterparts. However, their line profiles differ as simulated distributions drop off steeply at lower energies instead of containing a distinct downshifted peak. This suggests that the nature of the additional trapping caused by annealing at room temperature is separate from the trapping we observed from proton irradiation and that which is modeled in our numerical simulations. The distinction between these types of traps is important to maintaining the resolution of our GeDs and warrants more extensive study beyond the scope of this work. 

\subsection{High-temperature Anneals}
\label{subsec:hightempanneals}

\begin{figure}[htbp]
    \centering
    \captionsetup{skip=0.1pt} 
    \begin{subfigure}[b]{0.66\textwidth}
        \centering
        \includegraphics[width=\linewidth]{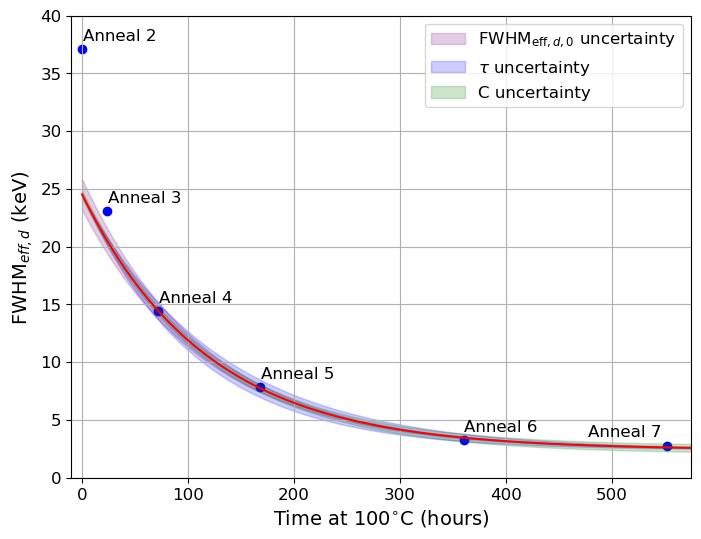}
        \caption{FWHM$_{\text{eff,d}}$ (component due to radiation damage) plotted against time spent annealing at $100^{\circ}$C, for hole-collecting strips utilizing a Cs-137 source. An exponential decay function is fit to the data points for anneals 4-7 and plotted in red. We find the best fit when FWHM$_{\text{eff}, d,0} = 22.1\pm 1.3$\,keV, $\tau = 118 \pm 12 $ hours, and $C = 2.4 \pm 0.3$\,keV. $\pm$1-sigma error margins for FWHM$_{\text{eff}, d,0}$, $\tau$, and C are shaded in purple, blue, and green, respectively.  }
        \label{fig:FWHM}
    \end{subfigure}
    
    \vspace{0.1cm}  

    \begin{subfigure}[b]{0.66\textwidth}
        \centering
        \includegraphics[width=\linewidth]{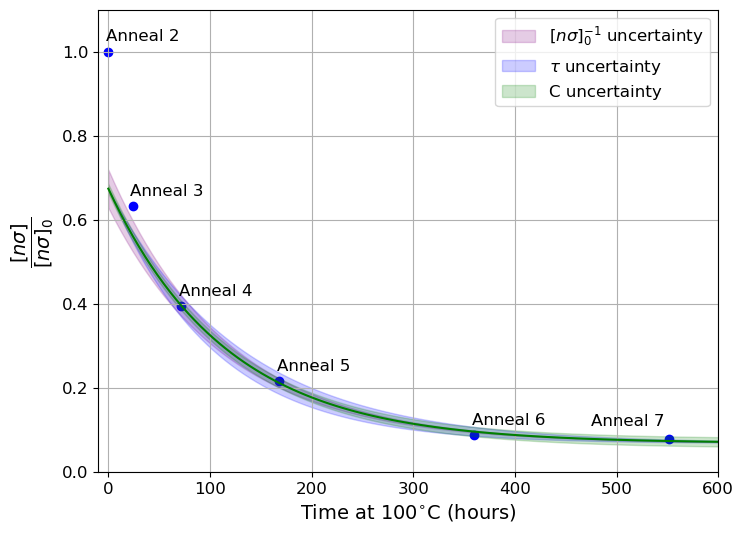}
        \caption{Decay of post-radiation trap density over time spent annealing at $100^{\circ}$C. The y-axis tracks the decrease in trap density (and increase in trapping product) by dividing the post-anneal 2 trapping product $[n\sigma]_0^{-1} = 37$ cm by the trapping product measured each successive anneal $[n\sigma]^{-1}$. An exponential decay function is fit to the data points for anneals 4-7 and plotted in green. We find the best fit when $[\frac{n}{n_0}]_0 = 0.61 \pm 0.04 $, $\tau = 117 \pm 14$ hours, and $C = 0.07 \pm 0.01$. $\pm$1-sigma error margins for $[\frac{n}{n_0}]_0$, $\tau$, and C are shaded in purple, blue, and green, respectively. }
        \label{fig:trapping_product}
    \end{subfigure}
    \vspace{0.05cm} 
    \caption{}
\end{figure}

After the first two anneals at $80^{\circ}$C, summing to 72 hours, we observed FWHM$_{\text{eff,h}}$ reduced to 37.2\,keV for hole-collecting strips. The photopeaks measured after each of the five $100^{\circ}$C anneals are plotted in Figure \ref{fig:anneals} and the repair of FWHM$_{\text{eff}, d}$ with time annealing at $100^{\circ}$C is marked by blue data points in Figure \ref{fig:FWHM}. We characterize a time scale for repair of primary damage at $100^{\circ}$C by fitting these data points to an exponential function plotted in red. Statistical errors on these data points are small ($<$\,0.2\,keV FWHM) compared to their scatter, so we assume the uncertainties on the fit parameters are dominated by systematic uncertainties, which were adjusted to achieve a reduced chi-squared value of unity. We omit data points from anneals 2 and 3 from this fit as their line profiles, pictured in Figure \ref{fig:simulatedpeaks}, show evidence of remaining secondary damage. The time constant best fit to the repair of primary damage is $118 \pm 12$ hours. To account for potential permanent damage which cannot be repaired by annealing at 100$^{\circ}$C, we included a constant term which is best fit at $2.4 \pm 0.3$\,keV. We apply the above time scale to the primary damage induced in this study and calculate a repair within 1\% of FWHM$_{\text{eff,h,d}}$ = 2.4\,keV in 490$\pm$30 hours of annealing at $100^{\circ}$C. 

 We use our simulation results (Section \ref{subsec:simulations}) to relate Cs-137 FWHM$_{\text{eff}}$ to the trapping product (Table \ref{table}) in order to track the way in which trap density changes with time annealing at $100^{\circ}$C. We plot the results in Figure \ref{fig:trapping_product}. The exponential decay fitted to these data, again omitting measurements from anneals 2 and 3, has a time constant of $117 \pm 14$ hours, a rate of repair consistent with that of FWHM$_{\text{eff,d}}$.


\section{Discussion}

\subsection{Primary vs. Secondary Defect Repair}
\label{subsec:80C}

 Our FWHM$_{\text{eff}}$ measurements and spectra collected after irradiation and then after room temperature cycling (detailed in Sections \ref{subsec:postrad} and \ref{subsec:roomtemp}, respectively) indicate that there are two distinct components to induced trapping, which we have identified as the primary and secondary components. This observation is confirmed by the study of secondary defects in \citet{Fourches1991} and \citet{Fourches_and_Walter}. Each type of defect will repair, under high temperature anneal, on a different timescale. Thus we model the repair of defects with time spent annealing as a double exponential function:
 \begin{align}
    FWHM_{\text{eff}, d} = A \cdot e^{-t/\tau_1} + B\cdot e^{-t/\tau_2} +C 
 \label{double_exp_eqn}  
 \end{align}
     
 We attempt to determine the two time constants at 80$^{\circ}$C for primary and secondary damage repair, $\tau_{1,80^{\circ}C}$ and  $\tau_{2,80^{\circ}C}$ . 

 We rely on our time constant derived for $100^{\circ}$C annealing in Section \ref{subsec:hightempanneals} to determine a repair time for primary damage at 80$^{\circ}$C. Annealing studies completed on radiation damaged p-type germanium have shown that the repair time of high-temperature annealing increases exponentially with decreasing temperature \cite{Brown1953, Waite1957}. In these studies, a 20$^{\circ}$C temperature decrease corresponds roughly to a repair time that is five times longer \cite{Brown1953}. Therefore, $\tau_{80^{\circ}C} = 5 \times \tau_{100^{\circ}C} =$ 590 hours.

We argue in Section \ref{subsec:hightempanneals} that the time constant obtained for annealing at  $100^{\circ}$C tracks the rate of repair for primary defects alone. Comparison of experimental and simulated line profiles in Figure \ref{fig:simulatedpeaks} reveals bimodal line profiles for anneals 2 and 3 characteristic of the distribution produced by room temperature cycling. By comparison, anneals 4-7 have single-peaked distributions which are in alignment with our simulations that model primary damage. This motivates us to omit measurements from anneals 2 and 3 from the fitted decay to obtain a repair of primary damage at $100^{\circ}$C. Furthermore the time scale we obtain from this fit (118 hours) is consistent with previous studies that report annealing times for repair of primary damage at $100^{\circ}$C on the scale of $\sim$ 100 hours \cite{Peplowski2019, Pehl1978}.

\begin{figure}
    \centering
    \includegraphics[width=0.75\linewidth]{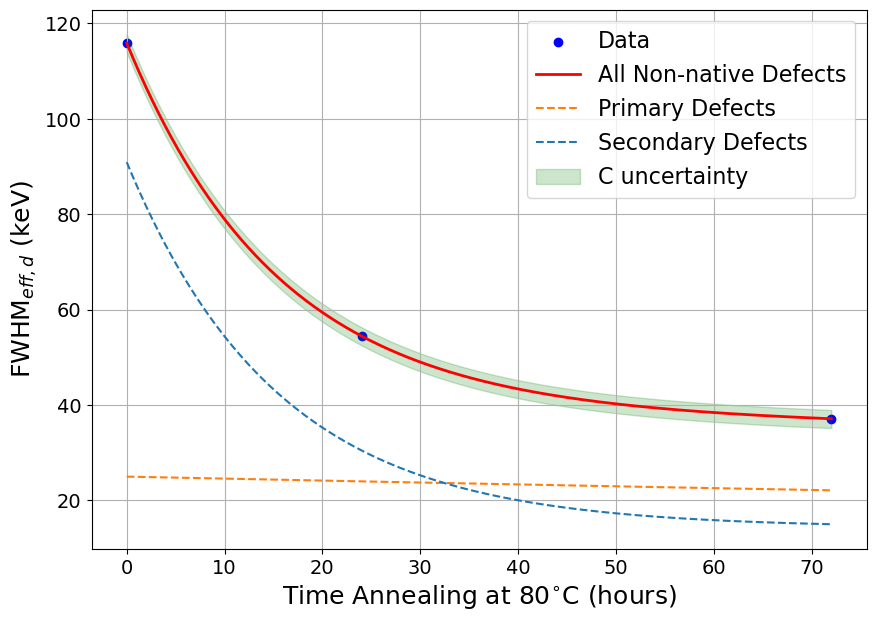}
    \caption{Anneals at $80^{\circ}$ fitted by a double exponential function plotted in red. Decay for primary defects is plotted in orange, while decay for secondary defects is plotted in blue. We assume $\tau_{1,80^{\circ}C} =$ 590 hours and A = 25$\pm$1.5\,keV. We find the best fit when  B  = 76.6$\pm$0.5\,keV, $\tau_{2,80^{\circ}C} = 15.4 \pm 0.1$ hours, and C = 14.3$\pm$1.9\,keV. Margins of error are obtained by propagating errors of the fixed parameters A and $\tau_{1,80^{\circ}C}$. $\pm$1-sigma error margins for C are plotted in green. }
    \label{fig:double_exp}
\end{figure}

To derive a rate for the repair of secondary damage at $80^{\circ}$C, we fit our measurements for FWHM$_{\text{eff}, d}$ for hole-collecting strips across $80^{\circ}$C anneals to Equation \ref{double_exp_eqn}. We fix $\tau_{1,80^{\circ}C} =$ 590$\pm$60 hours following the discussion above and A = 25$\pm$1.5\,keV in accordance with the amplitude of our fit in Figure \ref{fig:FWHM}. We plot data points in blue in Figure \ref{fig:double_exp} and attempt to fit decay of secondary damage, B (the amplitude of FWHM$_{\text{eff},d,sec}$), and a constant attributed to damage that cannot be repaired by $100^{\circ}$C annealing. The resulting decay of FWHM$_{\text{eff,d}}$, together with FWHM$_{\text{eff},d,pri}$ and FWHM$_{\text{eff},d,sec}$, is plotted in Figure \ref{fig:double_exp}. We obtain a time constant for the repair of secondary defects $\tau_{2,80^{\circ}C}= 15.4 \pm 0.1$ hours.

 This indicates that secondary defects, while detrimental to spectral performance, repair at lower temperatures and significantly faster than primary defects. This analysis appears consistent with the assumption that $80^{\circ}$C annealing procedures will not repair primary defects on a reasonable timescale for an instrument in space. 

 \subsection{Formation of Secondary Defects}
Additional questions remain about the nature of traps from secondary damage. Previous literature argues that these are deep traps in order to explain their effect on resolution \cite{Fourches_and_Walter}. However, the spectral peaks shifts inferred at separate detector depths from data collected after room temperature cycling diverge from charge trapping models outlined in \citet{Boggs_Pike_2023}, perhaps indicating detrapping effects. This would suggest, contrary to the conclusions of this earlier work, that secondary damage produces shallow traps which are repaired on a shorter timescale and at lower temperatures compared to primary damage, as demonstrated in Section \ref{subsec:80C}. Answering these questions is beyond the scope of this work, but could be key to maintaining the spectral resolution of GeDs operating in the space environment.

\subsection{Permanent Damage}
The remaining damage-induced spectral resolution after $>$\,550 hours of 100$^{\circ}$C annealing may be evidence of permanent damage. We spent 192 hours annealing at 100$^{\circ}$C between anneals 6 and 7 and observe only $\sim$\,0.3\,keV improvement in detector resolution. The measured FWHM$_{\text{eff}}$ for hole-collecting strips after anneal 7 remains $\sim$\,4.08\,keV with FWHM$_{\text{eff}, d}$ $\sim$\,2.79\,keV. This suggests that there is some small component of damage which our high-temperature annealing at 100$^{\circ}$C cannot eliminate. It is unclear whether this remainder is sustained from irradiation or cycling to room temperature and if it could be repaired by longer or higher temperature annealing procedures. A future study, in which irradiation and high-temperature annealing procedures are repeated without cycling the detector to room temperature, could answer these questions. 

\section{Conclusions}

We irradiated COSI GeDs with the maximum fluence expected over an extended $\sim$\,10 year mission, and induced additional damage by cycling the detector to room temperature, increasing the FWHM$_{\text{eff}}$ by $\sim$\,38 times is preradiation value. Through high-temperature annealing we are able to repair the spectral resolution to within 37\,$\%$ of its preradiation value, repairing to 4.08\,keV compared to the original 2.98\,keV. This notable recovery of detector resolution demonstrates that high-temperature annealing can repair even the most heavily damaged detectors. Furthermore, our measurements enable us to characterize a repair timescale for high-energy proton damage.

The high proton fluences used in this study, along with secondary damage from room temperature cycling, significantly overestimate the damage that COSI will receive during its primary mission. Guided by the linear relationship between fluence and measured FWHM$_{\text{eff}}$ demonstrated in Section \ref{subsec:postrad}, we calculate a linear regression and obtain the projected spectral resolution at the end of COSI's primary 2-year mission (FWHM$_{\text{eff,h,d}}$\,$\sim$\,8.8\,keV ). We apply the repair time scale derived in this work to estimate the anneal time needed to repair this resolution within 1\% of damage unrepairable at 100$^{\circ}$C (FWHM$_{\text{eff,h,d}}$ = 2.4\,keV). From these results, we anticipate that COSI spectral resolution can be repaired by annealing 100$^{\circ}$C in $320 \pm 30$ hours, on the scale of two weeks. Confirming this repair time by conducting irradiation and annealing studies at lower proton fluences will be the subject of future work.

\section{Acknowledgements}
This work was supported by the NASA Astrophysics Research and Analysis (APRA) program, grant 80NSSC22K1881.
The Compton Spectrometer and Imager is a NASA Explorer project led by the University of California, Berkeley with funding from NASA under contract 80GSFC21C0059.
This work was supported by the Laboratory Directed Research and Development Program of Lawrence Berkeley National Laboratory under U.S. Department of Energy Contract No. DE-AC02-05CH11231.
We would like to acknowledge the staff at the James M. Slater, MD, Proton Treatment and Research Center, Loma Linda University Medical Center beamline for their assistance in this study.

 \bibliographystyle{elsarticle-num-names}
 \bibliography{main}


\end{document}